\documentclass[pre,twocolumn,superscriptaddress,amsmath,amssymb,showpacs,noshowkeys,a4paper]{revtex4}
\usepackage{amsmath}
\usepackage{graphicx}
\usepackage{dcolumn}
\usepackage{bm}
\usepackage[usenames]{color}
\usepackage{epstopdf}
\definecolor{mygray}{gray}{0.5}

\newcommand{\afflangevin}{\address{Institut Langevin, ESPCI Paristech, CNRS - UMR 7587, PSL Research University, Universit\'e Pierre and Marie Curie,  1 rue Jussieu, 75005, Paris, France, EU}}
\newcommand{\affMSC}{\address{Laboratoire Mati\`{e}re et Syst\`{e}mes Complexes, Universit\'e Paris Diderot, Sorbonne Paris Cit\'{e}, CNRS - UMR 7057, 10 Rue A. Domon and L. Duquet, 75013 Paris, France, EU}}
\newcommand{\affMPQ}{\address{\emph{Present address:} Laboratoire Mat\'{e}riaux et Ph\'{e}nom\`{e}nes Quantiques, Universit\'e Paris Diderot, Sorbonne Paris Cit\'{e}, CNRS - UMR 7162, 10 Rue A. Domon and L. Duquet, 75013 Paris, France, EU}}
\newcommand{\affMIT}{\address{Department of Mathematics, Massachusetts Institute of Technology - 77 Massachusetts Avenue, Cambridge, MA 02139, USA}}

\newcommand\rbar{\overline{r}}
\begin{document}
\title{Polygonal instabilities on interfacial vorticities}
\author{M. Labousse}
\afflangevin
\affMSC
\affMPQ
\author{J.~W.M Bush}
\email{bush@math.mit.edu}
\affMIT
\begin{abstract}
We report the results of a theoretical investigation of the stability of a toroidal vortex bound by an interface. 
Two distinct instability mechanisms are identified that rely on, respectively, surface tension and fluid inertia, either of 
which may prompt the transformation from a circular to a polygonal torus. 
Our results are discussed in the context of three experiments, 
a toroidal vortex ring, the hydraulic jump,
and the hydraulic bump.
\end{abstract}

\maketitle

\section{Introduction}
Polygonal instabilities have been observed and reported in a variety of hydrodynamic systems across a wide range of scales~\cite{Perrard,EllegaardNature,Bohr99,Bush2,Labousse13PoF},~\cite{Bergmann,Couder83,Moisy04}, ~\cite{Widnall77,Maxworthy77,Saffman78},~\cite{Schubert},~\cite{Saturne1,Saturne2,Saturne3}. In many cases, the mechanism of instability remains poorly understood. We here 
shed some light on this class of problems by considering the instability of a fluid torus bound by an interface.

One of the most striking examples of polygonal instability is that of the hydraulic jump, as discovered by Ellegaard \textit{et al.}~\cite{EllegaardNature}. Usually, when a vertical jet of fluid strikes a horizontal plate, the flow spreads radially and a circular 
hydraulic jump arises at a critical radius~\cite{Rayleigh14,Tani49,Watson,Bush1,Kasimov08}. However, in certain parameter regimes, the 
axial symmetry is broken, leading to a polygonal jump (see Figure \ref{ressaut}a). The number of sides is strongly dependent 
of the fluid properties and the depth as well as the incoming flow rate~\cite{Bush2,Teymourtash}. Bohr \textit{et al.}~\cite{Bohr96,Bohr97} Andersen \textit{et al.}~\cite{Bohr10} and Watanabe \textit{et al.}~\cite{Bohr03} noted that a roller vortex downstream of the jump is a prerequisite for the formation of the polygonal pattern. Ellegaard \textit{et al.}~\cite{EllegaardNature} suggested that this polygonal transition may be induced by a weak line tension associated with the vortex that acts to minimize the circumference.  Bush \textit{et al.}~\cite{Bush2} and Teymourtash and Mokhlesi~\cite{Teymourtash} have investigated this system across a wide range of Reynolds and Weber numbers, and highlighted the critical role of surface tension. 
Indeed,  Bush \textit{et al.}~\cite{Bush2} reported that the addition of surfactant can suppress the polygonal instability entirely. 
The authors suggested that the instability may be due to a Rayleigh-Plateau like instability of the inner 
surface of the jump. This suggestion was pursued by Martens \textit{et al.}~\cite{Bohr12}, who developed a nonlinear model for the instability and successfully applied it, but did not consider the role of the roller vortex in the pressure distribution.   
Taken collectively, these studies suggest that surface tension and the roller vortex both play a crucial role in
the polygonal instability. \\
\\
Labousse and Bush~\cite{Labousse13PoF} reported that below a critical incoming flow rate, a plunging jet can give 
rise to a surface deflection called the hydraulic bump. The flow is marked by a subsurface poloidal vortex that is circular 
at low flow rates, but may destabilize into a polygonal form (see Figure \ref{bump}a). Owing to the relatively modest 
surface signature of the vortex, the structure is termed the hydraulic bump. We note that polygonal hydraulic bumps can 
also be observed in the presence of the hydraulic jump, presumably owing to the instability of the roller vortex downstream 
of the jump \cite{Bohr03,Bush2}. One may thus obtain polygonal jumps bound by polygonal bumps~\cite{Labousse13PoF} (e.g. see the six-sided outer surface structure in Figure~\ref{ressaut}a).\\ 
\\
Another striking example of polygonal instabilities has been discovered by Perrard \textit{et al.}~\cite{Perrard}, and is illustrated in Figure \ref{leidenfrost}a.
A fluid torus is contained in a circular trench heated beyond the Leidenfrost threshold \cite{Quere13}: the fluid is thus
levitated on the substrate and heated vigorously from below, resulting in a vigorous poloidal motion. The resulting 
fluid form is unstable: symmetry-breaking instabilities give rise to a polygonal inner surface (see Figure \ref{leidenfrost}a). \\
\\
In all three of these systems, vorticity and surface tension would appear to be significant.
We proceed by developing a theoretical model that captures the physics common to each of these three
systems. We first introduce the theoretical framework in Sec.~\ref{theoreticalframework}. Then we evaluate the linear stability of a fluid torus in Sec.~\ref{polygonalinstabilities}, the problem being an extension of Rayleigh-Plateau 
to the case of a toroidal geometry and an associated poloidal vortex (Figure 1). Two distinct stability mechanisms 
are identified in Sec.~\ref{instabilitymechanism} that rely on, respectively, surface tension and the poloidal swirl, and simple scaling laws are proposed 
to quantify the relative importance of these two destabilizing effects.  When possible, the results are compared to
previously reported data~\cite{Perrard,Bohr99,Bush2,Labousse13PoF} in Sec.~\ref{comparison}.
\section{Theoretical framework\label{theoreticalframework}}
\subsection{System parameters and dimensionless groups}
We consider a fluid torus with radii $R$ and $a$, density $\rho$, viscosity $\eta$ and surface tension $\gamma$. The main geometrical features of a torus are summarized in Appendix~\ref{appendixA}. As sketched in Figure \ref{tore}, the motion of the torus is defined by a poloidal swirling motion
\begin{equation}
\bm{\omega}=\omega \mathbf{e}_{\varphi}
\end{equation}
the latter being referred to as the poloidal vorticity. The local Reynolds number
\begin{equation}
Re=\rho\omega a^2/\eta
\end{equation}
is assumed to be sufficiently large that the effect of viscosity is negligible (see Table \ref{Parameter}). The relative 
magnitude of surface tension and inertia is prescribed by the Weber number, defined as
\begin{equation}
We=\dfrac{\rho \omega ^2 a^3}{\gamma}.
\end{equation} 
We neglect the effect of gravity. The dimensionless radius is defined by $\overline{r}=r/ R$, the aspect ratio of the torus by \begin{equation}
\chi=a/ R
\end{equation}
and the dimensionless distance from the $z$-axis  by 
\begin{equation}
\beta = \beta\left( \overline{r},\theta \right)= 1+\overline{r}\cos \theta.
\end{equation}
$\chi$ and $\rbar$ are taken to be small ($0<\rbar \leq \chi<0.3$). For the sake of simplicity, we consider a torus with circular section. The key system parameters and dimensionless groups are summarized in Table \ref{Parameter}.
\begin{table*}
\caption{Typical parameters and dimensionless numbers}
\label{Parameter}
\begin{center}
\begin{tabular}{lcccc}
Parameter & Notation& Polygonal torus& Polygonal jump & Polygonal bump \cr
\hline
density [kg/L]& $\rho$ &$\simeq$0.96& $\simeq$ 1.1 &$\simeq$ 1.1\\
Viscosity [cP]& $\nu$ &0.24 & 1-35  &60-70\\
Surface tension [mN.m$^{-1}$] & $\gamma$ & 58 & 60-70 & 68 \\
Vortex radius [cm] &$R$& $\simeq$ 2-3& $\simeq$ 1-4& $\simeq$ 2-5\\
Vortex radius [mm] &$a$& 4-8 & 1-10 & 2-4\\
\hline
Dimensionless group &  & &  & \cr
\hline
Aspect ratio &$\chi=a/R$& $\simeq 0.1$ & $0.1-0.2$ & $\simeq 0.05-0.08$\\
Local Weber number & $We=\rho \omega^2 a^3/\gamma$&$\sim 1-10$&$\sim 1-10$&$\sim 1- 10$\\
Local Reynolds number & $Re= \rho \omega a^2/\eta$&$\sim 100$&$\sim 10-150$&$\sim 10- 50$\\
\end{tabular}

\begin{tabular}{lcc}
\hline
\hline
Theoretical parameters & Notation& Range \cr
\hline
Toroidal coordinates& $(r,\theta,\phi)$ &$\left[0 ;a\right]\times \left[0 ;2\pi\right]^2$\\
Dimensionless toroidal coordinates& $(\rbar=r/R,\theta,\phi)$ &$\left[0 ;\chi \right]\times \left[0 ;2\pi\right]^2$\\
Distance to the $z$-axis & $R+r \cos\theta$ & $\left[R-a;R+a\right] $\\
Dimensionless distance to the $z$-axis & $\beta(\rbar,\theta)=1+\rbar \cos\theta$ & $\left[1-\chi;1+\chi\right] $\\
$\beta_\pi$&$\beta(\rbar=\chi,\theta=\pi)=(1-\chi)$&\\
Curvature & $C^{(0)}=\frac{1}{a}+\frac{\cos\theta}{R+a\cos\theta}$&\\
\end{tabular}
\end{center}
\end{table*}
\begin{figure}
\includegraphics[width =0.85 \columnwidth]{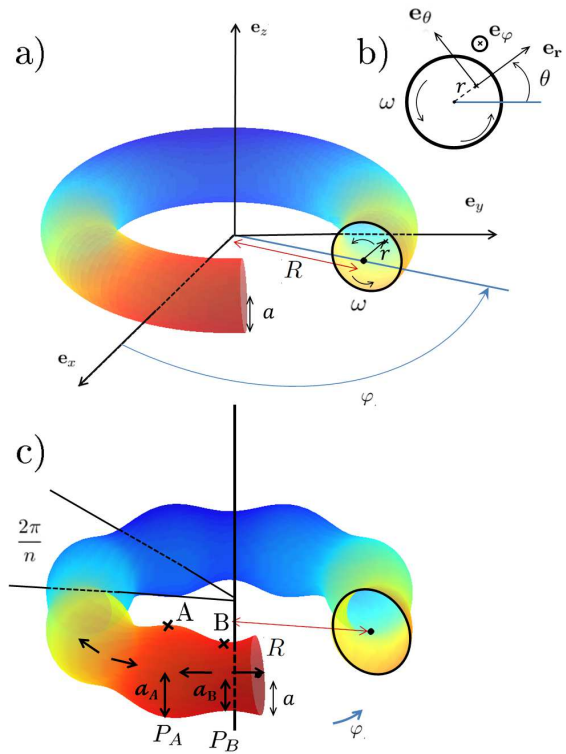}
\caption{a) The model system is described in terms of toroidal coordinates $r$, $\theta$, $\varphi$. There is a poloidal rotation with angular speed $\omega$. b) A section view of the torus of radius $a$ and the toroidal basis vectors. c) A schematic illustration of an octagonal instability.}
\label{tore}       
\end{figure}
\subsection{Operators in a toroidal basis}
The toroidal coordinates are expressible in terms of their cartesian counterparts $(X,Y,Z)$ in standard form:
\begin{equation}
\left\{
    \begin{array}{ll}     
        X=\left(R+r\cos{\theta}\right) \cos{\varphi }=R\beta \cos{\varphi }\\
        Y=\left(R+r\cos{\theta}\right) \sin{\varphi }=R\beta \sin{\varphi }     \\
        Z= r\sin{\theta}=R\overline{r}\sin{\theta}
     \end{array}
\right.
\end{equation}

The form of the differential operators in the toroidal coordinates $(r,\theta,\varphi)$, are summarized in Appendix~\ref{appendixA}. In the toroidal basis, $\left( \mathbf{e}_r,\mathbf{e}_{\theta },\mathbf{e}_{\varphi } \right)$, for an inviscid fluid, the Euler equations and the mass conservation can be written
\begin{small}
\begin{center}
\begin{equation}
\left\{
    \begin{array}{ll} 
\dfrac{D u}{D t}  -\dfrac{1}{R}\left[ \dfrac{v^2}{\rbar}+\dfrac{\cos{\theta } w^2}{\beta}\right] =-\dfrac{1}{R\rho }\dfrac{\partial P}{\partial \rbar}+\dfrac{f_r}{\rho}&(a) \\
        \dfrac{D v}{D t}  +\dfrac{1}{R}\left[ \dfrac{v\, u}{\rbar}+\dfrac{\sin{\theta } w^2}{\beta}\right] =-\dfrac{1}{R\rho }\dfrac{1}{\rbar}\dfrac{\partial P}{\partial \theta}+\dfrac{f_{\theta}}{\rho }&(b)\\
        \dfrac{D w}{D t}  +\dfrac{1}{R}\left[ \dfrac{w\, u \cos{\theta}}{\beta}-\dfrac{\sin{\theta } w\, v }{\beta}\right] =-\dfrac{1}{\rho }\dfrac{1}{R\beta}\dfrac{\partial P}{\partial \varphi}+ \dfrac{f_{\varphi}}{\rho}&(c)\\
        \dfrac{1}{\rbar\beta }\dfrac{\partial \rbar \beta u }{\partial \rbar}+\dfrac{1}{\rbar\beta}\dfrac{\partial \beta v}{\partial \theta}+\dfrac{1}{\beta}\dfrac{\partial w}{\partial \varphi}=0&(d)
         \end{array}
\right.
\label{Euler}
\end{equation}
\end{center}
\end{small}
with $P$ the pressure, $\mathbf{v}=\left(u,v,w \right) $ the velocity, $\rho$ the density and $D/Dt=\partial/\partial t+\mathbf{v}.\bm{\nabla}$. We denote by $\mathbf{f}$ the additional force field required for the basic state to be in equilibrium. 
\section{Polygonal instabilities\label{polygonalinstabilities}}

We proceed by analyzing the stability of the fluid torus. In subsection~\ref{steadystate}, we define the steady state and the mechanical balance. In subsection~\ref{stability} we analyze the linear stability of the torus. We discuss two asymptotic limits of the model in a last subsection~\ref{limits}.
\subsection{Steady State~\label{steadystate}}
 By way of capturing the essential common feature of the three flows of interest, we choose a purely poloidal vortical 
flow for the basic state velocity field $\mathbf{v}$. It is oriented along $\mathbf{e}_{\theta}$ and so may be expressed as
$\mathbf{v}=v\mathbf{e}_{\theta}$. The continuity equation imposes the condition
\begin{equation}
\dfrac{\partial \beta v}{\partial \rbar}=0
\end{equation}
which dictates that $v= \mathcal{F}(\rbar )/\beta$. 
The final form of $\mathcal{F}$ is found by prescribing a constant vorticity along $\mathbf{e}_{\varphi}$, specifically $\bm{\omega}=\omega \mathbf{e}_{\varphi}=(1/2)\bm{\mathrm{rot}}\mathbf{v}$, which yields $\mathcal{F}=\omega r=\omega \rbar R$. Consequently, the basic state velocity field may be expressed as
\begin{equation}
\mathbf{v}= \dfrac{\omega r}{\beta}  \mathbf{e}_{\theta}=\dfrac{\omega \rbar}{\beta}R  \mathbf{e}_{\theta},
\label{speedsteady}
\end{equation}
This steady state flow corresponds to a solid body rotation in a toroidal geometry, the simplest form that captures the essential
features of the three systems of interest.

The total curvature of torus is the sum of the poloidal and azimuthal contributions:
\begin{equation}
C^{(0)}=\bm{\nabla}.\mathbf{e}_r=C^{(0)}_{\theta}+C^{(0)}_{\varphi}
\end{equation} 
with
\begin{equation}
\left\{
    \begin{array}{ll} 
    C^{(0)}_{\theta}=\dfrac{1}{a}=\dfrac{1}{R\chi}\\
    \\
    C^{(0)}_{\varphi}=\dfrac{\cos\theta}{R+a\cos\theta}=\dfrac{1}{R}\dfrac{\cos\theta}{\beta(\rbar=\chi,\theta)}
         \end{array}
\right.
\label{courbure0}
\end{equation}

If unbalanced by external forces, surface tension will cause the torus to collapse into a sphere~\cite{Texier}. 
We note that in the three physical systems of interest, the radial force resisting this collapse has different origins. 
For example, in the Leidenfrost torus, the resisting radial force originates in the topography \cite{Perrard}. 
We here consider a body force density of the form
\begin{equation}
\bm{f}=\left(f_r(\rbar,\theta),f_{\theta}(\rbar,\theta), 0 \right).
\end{equation}
This force is required to maintain the toroidal shape of the ring, simultaneously exerting a radial force that
resists collapse, and satisfying the normal stress boundary condition on the toroidal surface. It
must thus satisfy the following  relations 
\begin{equation}
\left\{
    \begin{array}{ll}
    \displaystyle R\int\limits_{0}^{\rbar}f_r(\rbar,\theta) d\rbar=\dfrac{\gamma}{R}\left(\dfrac{1}{\chi}+\dfrac{\cos \theta}{\beta(\rbar,\theta)} \right)\\
    \displaystyle R\int\limits_{0}^{\theta}\rbar f_{\theta}(\rbar,\tilde{\theta}) d\tilde{\theta}=  \dfrac{\gamma}{R}\left(\dfrac{1}{\chi}+\dfrac{\cos \theta}{\beta(\rbar,\theta)} \right)
    \end{array}
\right.
\label{conditionbodyforce}
\end{equation}
in the bulk to meet the boundary conditions. \\

The steady form of the governing set of Equations~\ref{Euler} can be expressed as:
\begin{equation}
\left\{
    \begin{array}{ll}
    
         \dfrac{\partial P}{\partial \overline{r}}= \rho \omega ^2 R^2 \dfrac{\overline{r}}{\beta^2}+  R f_r\\
       \dfrac{\partial P}{\partial \theta}  = R\rbar f_{\theta}\\ 
       \dfrac{\partial P}{\partial \varphi}=0
    \end{array}
\right.
\label{Eulersteady}
\end{equation}
The aspect ratio of the torus remains small, so Equations~\ref{Euler} can be expressed to leading order in $\rbar$ as detailed in Appendix~\ref{Appendixsupplementaire}. This set of equations can be integrated, using Equations~\ref{conditionbodyforce}, to yield
\begin{equation}
\begin{array}{rl}
P(\rbar,\theta)=&P_{0}+\rho \dfrac{\omega ^2 R^2}{2}(\overline{r}^2-\chi^2) \\
&+\dfrac{\gamma}{R}\left( \dfrac{1}{\chi}+\dfrac{\cos{\theta}}{\beta(\rbar,\theta)}\right) +\mathcal{O} \left( \overline{r}^3 \right).
\end{array}
\label{Equationpression}
\end{equation}
with $P_0$ being a constant pressure. The resulting pressure $P$ can be simply seen as resulting from the combined effect of inertia and surface tension. 
Note that equation~\ref{conditionbodyforce} insures that the normal stress condition 
\begin{equation}
P(\rbar=\chi,\theta) -P_0=\gamma C^{(0)}
\end{equation} 
is satisfied: the Laplace pressure corresponds to that of a liquid torus with a local curvature $C^{(0)}$. 
\subsection{Stability~\label{stability}}
The perturbations of the steady state in pressure $\tilde{p}$ and velocity vector $\bm{\varepsilon}=(\varepsilon_r/\beta,\varepsilon_{\theta}/\beta,\varepsilon_{\varphi})$ are defined through
\begin{equation}
\left\{
    \begin{array}{l}
      \mathbf{V}_{\mathrm{total}}=\mathbf{v}+
      \begin{pmatrix}
      \dfrac{\varepsilon_r(\rbar)}{\beta } \\
      \dfrac{\varepsilon_{\theta}(\rbar)}{\beta } \\
\varepsilon_{\varphi}(\rbar)
\end{pmatrix}
=\mathbf{v}+
\begin{pmatrix}
\dfrac{\varepsilon_{r,0}(\rbar)}{\beta } \\
\dfrac{\varepsilon_{\theta ,0}(\rbar)}{\beta } \\
\varepsilon_{\varphi ,0}(\rbar) \\
\end{pmatrix} e^{\sigma t}e^{in\varphi}\\ \\
        P_{\mathrm{total}}=P+\tilde{p}=P+\tilde{p}_0 e^{\sigma t}e^{in\varphi},\\     
    \end{array}.
\right.
\end{equation}
where $\sigma$ is the growth rate, and $n$ the number of sides of the associated polygonal form. We assume that $\tilde{p}\ll P$ and $\Vert \bm{\varepsilon}\Vert \ll \Vert \mathbf{v} \Vert $. The $1/\beta$ factor in the $r$ and $\theta$ components of $\bm{\varepsilon}$ can be simply seen as a trick to compute easily the first-order expansion of the conservation equation $\bm{\nabla}.\bm{\varepsilon}=0$. We restrict the class of perturbations to azimuthal modes and neglect the poloidal ones, in which case the disturbance amplitudes $(\varepsilon_{i,0})_i$ are independent of $\theta$.

By taking into account the Euler (Eqs~\ref{Euler}~(a)-(c)) and continuity equations (Eq.~\ref{Euler} (d)), a first-order expansion in $\bm{\varepsilon}$ and $\tilde{p}$ leads to 
 \begin{equation}
 \left\{
    \begin{array}{ll}
       A\varepsilon_r-B\varepsilon_{\theta}=-\dfrac{\partial \tilde{p}}{\partial \rbar}&(a)\\
        C\varepsilon_{\theta}+\varepsilon_{r}D=0&(b)\\
         E\varepsilon_{\varphi}=-in\tilde{p}&(c)\\
         \dfrac{1}{\rbar}\dfrac{\partial \rbar \varepsilon_r}{\partial \rbar}+in\varepsilon_{\varphi}=0 . &(d)
    \end{array}
\right.
\label{equationlinearise}
\end{equation}
All the terms $A, B, C, D, E$, depending on $\rbar$ and $\beta$, are detailed in Appendix~\ref{appendixB}.
The set of equations~\ref{equationlinearise} gives  
 \begin{equation}
 \left\{
    \begin{array}{ll}
       \varepsilon_r=-F\dfrac{\partial \tilde{p}}{\partial \rbar}&(a)\\
        \varepsilon_{\theta}=-\dfrac{D}{C}\varepsilon_{r}&(b)\\
         \varepsilon_{\varphi}=-\dfrac{in}{E}\tilde{p}&(c)\\
         \dfrac{1}{\rbar}\dfrac{\partial \left(\rbar F\dfrac{\partial \tilde{p}}{\partial \rbar}\right)}{\partial \rbar}-\dfrac{n^2}{E}\tilde{p}=0 . &(d)
    \end{array}
\right.
\label{equationlinearise3}
\end{equation}
with $F=C/(AC+BD)$. We restrict our angular parameter to $\theta=\pi$ which corresponds approximately to the angle at which polygonal patterns are observed in the three experiments of interest. Evaluated at angle $\theta=\pi$, Eq.~\ref{equationlinearise3}-d leads to a second order equation in $\tilde{p}$
\begin{equation}
\rbar^2 \dfrac{\partial^2\tilde{p}}{\partial\rbar^2}+\rbar \dfrac{\partial\tilde{p}}{\partial\rbar}-\overline{r}^2\tilde{n}^2 \tilde{p}+\mathcal{O} \left( \rbar^4\right) =0,
\label{equadiff}
\end{equation}
with 
\begin{equation}
\tilde{n}=n\sqrt{1+4\left( \dfrac{\omega}{\sigma}\right) ^2}.
\end{equation}
The passage from Eq.~\ref{equationlinearise3}-d to Eq.~\ref{equadiff} is detailed in Appendix~\ref{appendixC}. Note that Eq.~\ref{equadiff} has been evaluated at $\theta=\pi$ for the sake of simplicity but could be extended for any poloidal angle $\theta$.

An analytical solution of (\ref{equadiff})  can be computed by using power series. 
A second order expansion in $\rbar$ leads us to 
\begin{equation}
\tilde{p}  =\xi_0.I_0\left(\tilde{n}\rbar \right) +\mathcal{O} \left( \rbar^3\right)  
\label{expressionpressionlinearise}
\end{equation}
with $\xi_0$ constant and $I_{\nu}$ the modified Bessel function of the first kind of order $\nu$. \\

Determining the growth rate of the mode $n$ as a function of the control parameters requires considering the boundary conditions. We denote by 
\begin{equation}
\begin{array}{ll}
H(\rbar,\theta,t)&=(\rbar-\chi)-\int_{0}^t dt\;\bm{\varepsilon}.\bm{e}_r\\
&=(\rbar-\chi)-\dfrac{\varepsilon_{r}(\rbar,t)-\varepsilon_{r}(\rbar,0)}{R\beta\sigma}
\end{array}
\end{equation}
the surface functional with $H=0$ on the perturbed surface. We denote $\varepsilon_{\chi}=\varepsilon_{r,0}\left( \rbar=\chi\right) $. The curvature of the perturbed surface is given by the divergence of its normal vector $\bm{n}=(\bm{\nabla }H)/\Vert \bm{\nabla }H \Vert$, specifically\\
 \begin{equation} C= \bm{\nabla} \cdot \bm{n} = C^{(0)}+\dfrac{\varepsilon _{\chi}}{\sigma\beta} C^{(1)},
 \end{equation}
  with\\
 \begin{equation}
 \left\{
    \begin{array}{l}
        C^{(0)}= \dfrac{1}{R}\left( \dfrac{1}{\chi} +\dfrac{\cos{\theta}}{\beta\left(\rbar=\chi,\theta \right)} \right) \\
       C^{(1)}=\dfrac{-1}{R}\left(\dfrac{\cos{\theta}}{\chi\beta\left(\chi,\theta \right)}+\left(\dfrac{\sin{\theta}}{\beta\left(\chi,\theta \right)}\right)^2-\left( \dfrac{n}{\beta\left(\chi,\theta \right)} \right)^2\right)       
    \end{array}
\right.
\end{equation}
The boundary conditions link $\varepsilon _{\chi}=\varepsilon _{\rbar,0}(\rbar=\chi)$ with $\tilde{p}(\rbar=\chi,\theta=\pi)$ and its derivative $\partial_{\rbar}\tilde{p}(\rbar=\chi,\theta=\pi)$ as follows
\begin{equation}
\left\{
    \begin{array}{ll}
    \dfrac{\varepsilon _{\chi} \sigma}{\beta_{\pi}} \left( 1+ 4K\dfrac{\omega ^2}{\sigma ^2}\right) =-\dfrac{1}{R\rho} \left( \dfrac{\partial \tilde{p}}{\partial \rbar}\right)_{\rbar=\chi,\theta=\pi}&(a)\\
P+\tilde{p}=\gamma \left( C^{(0)} +\dfrac{\varepsilon _{\chi}}{\beta_{\pi} \sigma} C^{(1)}\right)&(b)\\
\end{array},
\right.
\label{boundaryconditions}
\end{equation}
with $K=1+5\chi/2+9\chi^2/2+\mathcal{O}\left( \chi^3\right)$. The origin of $K$ is given in Appendix~\ref{appendixC}. Note that Eq.~\ref{boundaryconditions}(a) arises from the combination of the linearised Eqs~\ref{equationlinearise}(a) and \ref{equationlinearise}(b).  Eq.~\ref{boundaryconditions}~(b) is the pressure boundary condition. Combining Eqs~\ref{boundaryconditions}(a) and~\ref{boundaryconditions}(b) with Eq.~\ref{expressionpressionlinearise} gives a relation for the growth rate $\sigma$ as a function of the control parameters and the mode number $n$. This relation takes the form 
\begin{equation}
\begin{array}{rl}
\left( \dfrac{\sigma}{\omega}\right)^2 \sqrt{1+4K\left( \dfrac{\omega}{\sigma}\right)^2 } =& \dfrac{\chi n}{1-\chi} \dfrac{I_1 \left( \tilde{n}\chi\right) }{I_0 \left( \tilde{n}\chi\right)} \lbrace 1+\\
& \dfrac{1}{We} \left[1-\mathcal{C}_{\chi}  \right] \rbrace .
\end{array}
\label{equation}
\end{equation}
with $\mathcal{C}_{\chi}=\chi/(1-\chi)+(n\chi)^2/(1-\chi)^2$. The term  $(1-\mathcal{C}_{\chi})/We$ denotes the dimensionless surface tension contribution, and the constant $1$ is the dimensionless signature of the poloidal vortex. In accordance with the results of Hocking \textit{et al.}~\cite{Hocking59}, Ponstein~\cite{Ponstein59}, Pedley~\cite{Pedley1} 
and Kubitschek \textit{et al.}~\cite{Weidman06} for the case of a cylinder of fluid, the poloidal 
vorticity $\omega$ destabilizes the system. The vorticity also extends the range of unstable 
wavelengths below that of the standard Rayleigh-Plateau threshold. By approximating 
$I_1(\tilde{n}\chi)/I_0(\tilde{n}\chi)\simeq I_1(n \chi)/ I_0(n\chi)$, as is valid provided $\omega\ll \sigma$, 
the maximum growth rate is found numerically. For a given Weber number $We$, the maximum 
of the real part of the growth rate $\sigma^2$ and the corresponding $n$ are found.

\begin{table*}
\caption{Adaptation of the predicted growth rate of polygonal instability to the three experimental cases of interest: 
(a) the Leidenfrost torus,  (b) the hydraulic jump and (c) the hydraulic bump. The notations and geometry are 
specified in Figures~\ref{leidenfrost}-\ref{bump}.}
\label{Growth}
\begin{center}
\begin{tabular}{c|l|l|l|l}
 &Case & Growth rate & Aspect ratio $\chi$ & Weber number $We$\cr
\hline
(a)&Tor. Leid.& \begin{footnotesize}
$\dfrac{\sigma^2}{\omega^2} \sqrt{ 1+4K\left( \dfrac{\omega}{\sigma}\right)^2 } \simeq \dfrac{\chi n}{1-\chi} \dfrac{I_1 \left( \tilde{n}\chi\right) }{I_0 \left( \tilde{n}\chi\right)} \lbrace 1+\dfrac{1}{We} \left(1- \mathcal{C}_{\chi} \right)\rbrace$
\end{footnotesize} &$\chi=\dfrac{a}{R_{int}+a}$ & $We=\dfrac{\rho v_{\pi}^2 a}{\gamma}$ \cr
(b)&Hyd. jump& \begin{footnotesize}$\dfrac{\sigma^2}{\omega^2} \sqrt{ 1+4K\left( \dfrac{\omega}{\sigma}\right)^2 } \simeq   \dfrac{\chi n}{1-\chi} \dfrac{I_1 \left( \tilde{n}\chi\right) }{I_0 \left( \tilde{n}\chi\right)} \lbrace 2+ 
\dfrac{1}{2We} \left(1- \mathcal{C}_{\chi}  \right) \rbrace$ \end{footnotesize} &$\chi=\dfrac{a}{r_j+a}$ &$We=\dfrac{\rho \omega^2 a^3}{\gamma}$\cr
(c)&Hyd. bump& \begin{footnotesize}
$\dfrac{\sigma^2}{\omega^2} \sqrt{ 1+4K\left( \dfrac{\omega}{\sigma}\right)^2 } \simeq \dfrac{\chi n}{1-\chi} \dfrac{I_1 \left( \tilde{n}\chi\right) }{I_0 \left( \tilde{n}\chi\right)} \lbrace 1+\dfrac{1}{We} \left(1- \mathcal{C}_{\chi } \right)\rbrace$
\end{footnotesize}& $\chi\sim \dfrac{H/2}{r_b}$&$We=\dfrac{\rho \omega^2 a^3}{\gamma}$\cr
\end{tabular}
\end{center}
\end{table*}
\subsection{Rayleigh-Plateau and Ponstein/Hocking/Pedley limits~\label{limits}}

For $We\ll 1$, the standard Rayleigh-Plateau (indexed as R-P) instability is recovered. Indeed, taking the limit of a cylinder, $\chi \rightarrow 0$, and keeping the product $n\chi$ constant, yields
\begin{equation}
\sigma_{\mathrm{R-P}} ^2=  ka\dfrac{I_1(ka)}{I_0(ka)} \dfrac{\gamma}{\rho a^3} \left( 1-(ka)^2 \right) +\mathcal{O} \left( \chi  \right).  
\end{equation}
where $k$ is the wave number of the disturbance given by $k=n\chi/a$. This result corresponds precisely to the relation found by 
Rayleigh~\cite{Rayleigh79}. Moreover, for $We\ll 1$, Rayleigh's instability criteria indicates that $ka\sim 1$, that is, $n\chi\sim 1$ 
in the present case. \\

In the limit of the cylindrical case and $We\gg 1$, one can replace $n\chi$ in the growth rate equation~\ref{equation} by $k a$, which gives 
\begin{equation}
\dfrac{\sigma_{\mathrm{P}}^2}{\omega ^2}(1+4\dfrac{\omega ^2}{\sigma_{\mathrm{P}}^2})=\dfrac{k'aI_1(k'a)}{I_0(k'a)}\left(1+\dfrac{1}{We}(1-(ka)^2)\right),
\end{equation} 
with $k'^2=k^2( 1+4(\omega/\sigma_{\mathrm{P}})^2)$ and thus we recover the results of Ponstein~\cite{Ponstein59} (indexed P), Hocking and Michael ~\cite{Hocking59} 
and Pedley~\cite{Pedley1}. In this regime, the most unstable mode is given by~\cite{Pedley1}  
\begin{equation}
(ka)^2\simeq \dfrac{1+We}{3}.
\end{equation} 
We note that for small aspect ratios ($\chi<0.1$), our predictions 
are closer to the cylindrical case, as one expects. 
\section{Instability mechanism\label{instabilitymechanism}}
\subsection{Scaling laws}
The polygonal shape arises from the combined destabilizing influences of the surface tension and the poloidal vortex. Imagine a perturbation to the torus giving rise to constricted 
and expanded regions near points B and A, respectively (see Figure \ref{tore}c).\\

In the capillary regime $We\ll 1$, surface tension dominates inertial terms and we recover the standard Rayleigh-Plateau instability. The mechanism is associated with the difference of the Laplace pressure between the constricted and expanded regions. One of the principal radii of curvature is positive in the zone A, and negative in the zone B. 
The resulting pressure difference between these two points drives
flow away from the constriction, thus amplifying the initial perturbation. The presence of the poloidal vorticity may likewise
prompt instability. \\

In the inertial regime $We\gg 1$, the dynamic pressure difference $d P_v$ dominates the Laplace pressure. The conservation of circulation $\Gamma$ requires that $2\pi va=\Gamma$, which indicates that the variation of speed squared $dv^2$ depends on the variation in radius $\delta$ as $dv^2 \sim \Gamma^2 \delta /a^3$. Also, the dynamic pressure difference scales as $d P_v\sim \rho \Gamma^2 \delta /a^3$ while the difference of curvature pressure can be expressed as $\gamma \delta /\lambda^2$. The balance of these two pressure differences yields $\gamma/\lambda^2 \sim \rho \omega ^2 a $ \textit{ i.e.} $\left(a/\lambda\right)^2\sim \rho \omega ^2 a^3/\gamma$, from which it follows that 
$n\chi\simeq ka \sim \sqrt{We}$, in accordance with the results of Pedley~\cite{Pedley1}.\\
\subsection{Effect of the asymmetry}
The system is further destabilized by the toroidal geometry, specifically by the asymmetry between the inner $(\theta=\pi )$ and the outer $(\theta=0)$ sides of the torus. In the capillary regime ($We\ll 1$), the difference of curvature pressure 
\begin{equation}
\begin{array}{ll}
\gamma \left( C^{(0)}(\theta=\pi)- C^{(0)}(\theta=0)\right)&=-\dfrac{\gamma}{R} \left(\dfrac{1}{1-\chi}+\dfrac{1}{1+\chi}\right)\\
 &=-\dfrac{2\gamma/R}{(1-\chi^2)}\\
 &<0
\end{array}
\end{equation} 
imposes a pressure difference that tends to straighten out the roller locally. Similarly, in the inertial regime ($We\gg 1$), the difference of Bernoulli pressure on the inner $(\theta=\pi)$ and outer $(\theta=0)$ sides of the vortex  
\begin{equation}
\begin{array}{ll}
\rho\left(v^2_{\theta}(\theta=\pi)- v^2_{\theta}(\theta=0)\right)&=\rho \omega^2 a^2\left(\dfrac{1}{(1-\chi)^2}-\dfrac{1}{(1+\chi)^2}\right)\\
&=4\rho \omega a^2\dfrac{\chi}{(1-\chi^2)^2}\\
&>0
\end{array}
\end{equation}
does likewise. These analogous local tendencies towards a straight vortex, in conjunction with the global topological constraint associated with toroidal geometry, may lead to a piecewise straight configuration \textit{i.e.} a polygonal pattern.  

In what follows, we apply the theoretical developments of section~\ref{polygonalinstabilities} to the three different experiments of interest.
\section{Comparison with three related physical systems \label{comparison}}
\subsection{The Leidenfrost torus\label{theleidenfrosttorus}}
\begin{figure}
\resizebox{0.5\textwidth}{!}{%
  \includegraphics{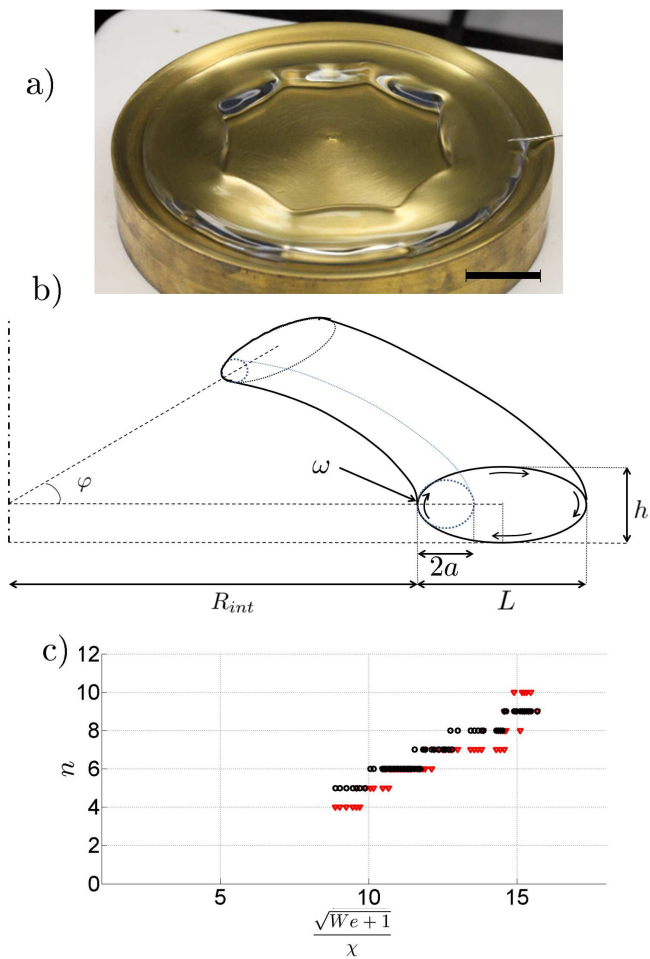}
}
\caption{a) The polygonal Leidenfrost torus \cite{Perrard}. A torus is fixed in a hot toroidal channel that induces  
poloidal vorticity within the core. Evaporation is compensated for by injection of fluid.  The inner
surface changes from circular to polygonal. Scale bar, $2$ cm. b) A schematic defines the principal geometrical features used 
in our theoretical model. We schematize in dashed lines the osculating torus. Our theoretical steady state describes reasonably well the physical situation at the inner side of this torus.  c) The observed dependence of the number of polygonal sides $n$ on $\sqrt{We+1}/\chi$.\textcolor{red}{$\bigtriangledown$}  indicate the experimental data from Perrard \textit{et al.} \cite{Perrard}, and $\circ$, our predictions
for the most unstable mode. Fig~\ref{leidenfrost}a, is used with Permission from Perrard \textit{et al.}~\cite{Perrard}.}
\label{leidenfrost}       
\end{figure}
We proceed by considering the experimental investigation reported by Perrard \textit{et al.}~\cite{Perrard}. 
Experimentally, the real base shape is a torus with an ellipitical cross section as sketched in the figure~\ref{leidenfrost}b. 
Nevertheless, as the polygonal forms are confined to the inner surface of the vortex ($\theta\simeq \pi$), we consider 
the radius of curvature there. At the inner side of the osculating torus (see Fig.~\ref{leidenfrost}-b), the curvature and pressure distribution can locally be described by our idealized steady state. 
We denote by $L$ the major axis and $h$ the minor axis. Near $\theta=\pi$, and near 
the surface, the flow may be approximated by our general theoretical framework. $a$ is defined as the radius of the 
osculating torus and can be approximated as the semi-minor axis $h/2$. Moreover the torus is confined to a circular trough 
that accounts for the required counterforce.
Our extrapolation from the theoretical framework to this experimental case is described in Table \ref{Growth}a,
where the torus radius is evaluated as $a\simeq h/2$.\\

The number of sides $n$ corresponding to the maximum growth rate is found numerically from the dispersion relation (Equation a in Table \ref{Growth}) and by using $\chi$ and $We$ from experiments. To simplify the algebra, we make the approximation 
$I_1\left( \tilde{n}\chi \right)/I_0\left( \tilde{n}\chi \right)\simeq I_1\left( n\chi \right)/I_0\left( n\chi \right)$ in (12) which is valid
provided $\tilde{n}\chi \ll 1$. 
The growth rate will depend in general on $(n,\chi, We)$. 
However, for the cylindrical case, Pedley \cite{Pedley1} and Hocking and Michael \cite{Hocking59} demonstrate 
that a two-dimensional representation  $(n,\sqrt{(We+1)}/\chi)$ is suitable, which in our case remains  a good approximation provided the aspect ratio is small. 
Figure \ref{leidenfrost}c indicates the dependence of the number of polygonal sides $n$ on $\sqrt{We+1}/\chi$. 
For $n=5$ to $9$, the theory adequately collapses the experimental data.  
 
\subsection{The hydraulic jump\label{thehydraulicjump}}
\begin{figure}
\resizebox{0.5\textwidth}{!}{%
  \includegraphics{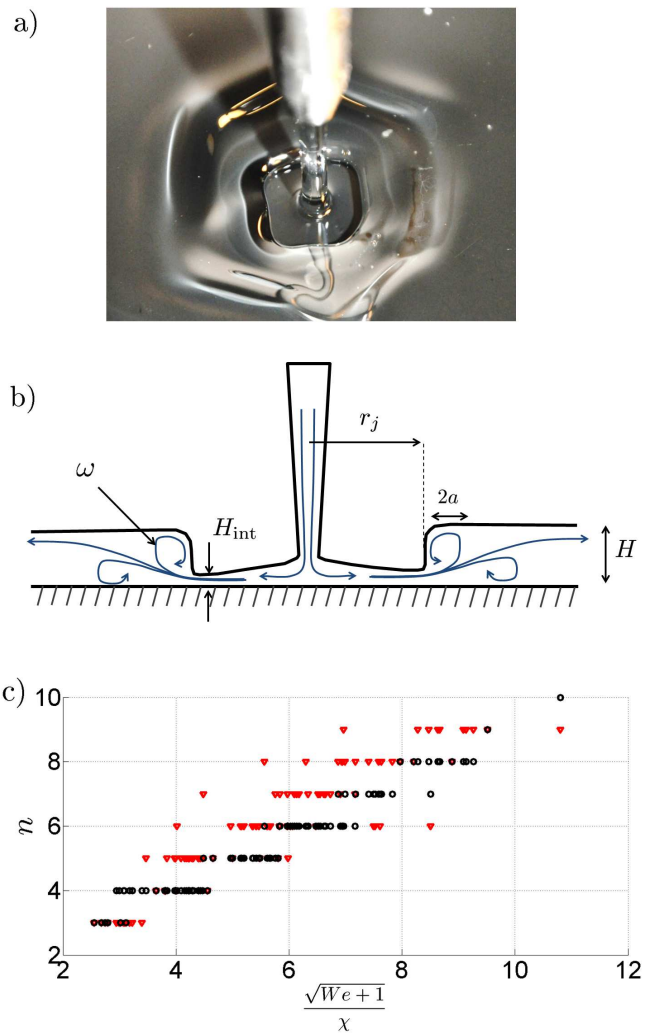}
}
\caption{a) A four-sided hydraulic jump within a six-sided hydraulic bump \cite{Labousse13PoF}. b) Schematics and notation for the polygonal hydraulic jump. c) The dependence of the number of polygonal sides $n$ on $\sqrt{We+1}$, where $We=\rho \omega^2 a^3/\gamma$. The experimental data \textcolor{red}{$ \bigtriangledown $} of Ellegaard \textit{et al.} \cite{Bohr99} are
plotted along with our theoretical predictions: $\circ$. We now note that the data of Teymourtash and Mokhlesi~\cite{Teymourtash} is consistent with that of Ellegaard et al.~\cite{Bohr99}}
\label{ressaut}       
\end{figure}
We next consider the geometry of the hydraulic jump, and assume a roller vortex just downstream of the jump.
As indicated on Figure \ref{ressaut}, we denote by $H_{\mathrm{int}}$ and $H$ the fluid depth, respectively up- and downstream 
of the jump, which has a radius $r_j$. The radius of the poloidal vortex ring $a$ can be approximated by 
$a\approx (H-H_{\mathrm{int}})/2$. The poloidal vorticity $\omega$ can be roughly deduced from mass conservation:
$a\omega \approx Q/(2\pi (H-2a) (R_j+a))$, where $Q$ is the total incoming flux.  
The dispersion relation is modified to account for the difference between the theoretical framework and the 
experimental configuration (Equation b in Table \ref{Growth}). First, surface tension only influences the inner surface of the roller vortex. We thus roughly approximate 
the curvature contribution $[1-\mathcal{C}_{\chi}]/We$ ((Eq.~\ref{equation})) by  $[1-\mathcal{C}_{\chi}]/(2We)$ ((Eq.~\ref{Growth}-b)). Second, the position of the jump and its associated vortex are determined by the incoming flow. We note 
that at the jump position, the Bernoulli pressure, $p\sim \rho v^2\sim \rho (\omega a)^2$  typically exceeds the radial shear 
stress  $\tau \sim \eta v/a$ by at least an order of magnitude, e.g. $p/\tau \sim \omega a^2/(\eta/\rho)=5\times 2\pi \times(4\times 10^{-3})^2/10^{-5}\sim 50$. Consequently, we add this incoming Bernoulli pressure term, 
i.e. $1$ in the corresponding dimensionless notation (Eq.~\ref{Growth}-b).
One then expects that the growth rate at $\theta=\pi $ can be written as described in Table \ref{Growth}b. 

Figure \ref{ressaut}c compares the experimental results from Ellegaard \textit{et al.}~\cite{Bohr99} with the theoretical predictions. 
The theoretical model agrees qualitatively with the data; however, the substantial scatter in the data 
precludes a strong conclusion. This scatter underscores the limitations of our model in describing this relatively 
complex fluid configuration. First, we note that we have neglected hydrostatic pressure, whose influence on the
polygonal jump has been demonstrated by Bush \textit{et al.} \cite{Bush2} and Martens \textit{et al.}~\cite{Bohr12}. Including the data set of Bush \textit{et al.} 
\cite{Bush2}, who explored a wider range of Weber and Bond number, only increases the scatter. 
Another limitation arises from the uncertainty on the radial extent of the roller vortex, and the associated 
uncertainty in the aspect ratio $\chi$, to which our model predictions are quite sensitive. 
Thus, while our simplified theoretical approach does capture some features of the polygonal jump 
instability, it also reaches its limits for this relatively complex configuration.

\subsection{The hydraulic bump\label{thehydraulicbump}}
\begin{figure}
\resizebox{0.5\textwidth}{!}{%
  \includegraphics{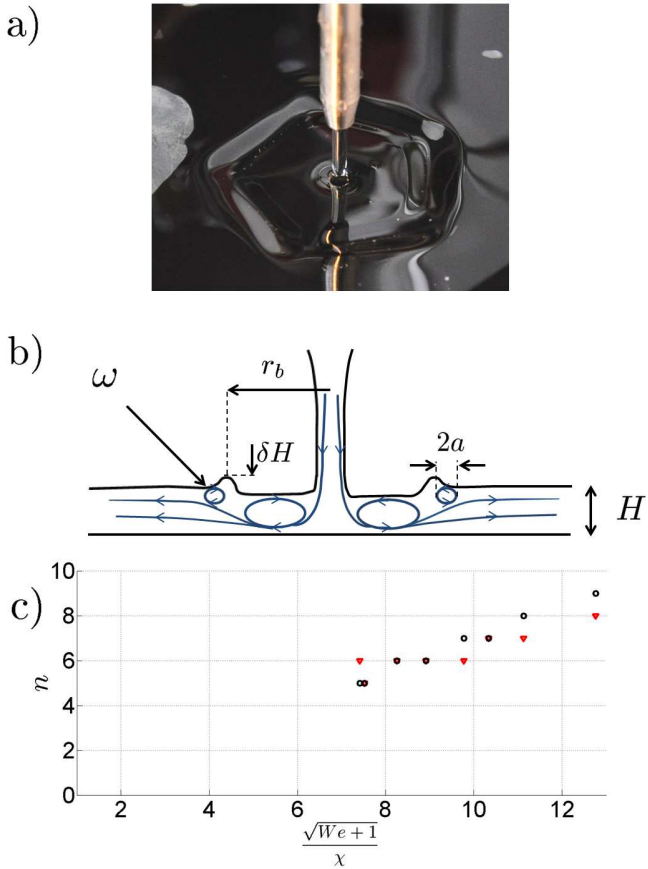}
}
\caption{a) Illustration of the five sided polygonal bump \cite{Labousse13PoF}. b) Schematic of the hydraulic bump. c) The dependence of the number of polygonal sides $n$ on $\sqrt{We+1}/\chi$. Data from Labousse and Bush~\cite{Labousse13PoF} \textcolor{red}{$\bigtriangledown$}
are presented with the results of our model predictions ($\circ$).}
\label{bump}       
\end{figure}

Given the relatively small surface signature of the hydraulic bump~\cite{Labousse13PoF}, we expect the 
subsurface vortex to be primarily responsible for the polygonal instability (see Figure \ref{bump}a). 
As sketched in Figure \ref{bump}b, we denote the bump radius by $r_b$, the bump height by $\delta H$ and the outer depth by $H$. 
The vortex ring has radius $a \sim \delta H$ with a poloidal vorticity $\omega$ that may be approximated as 
$ Q/(2\pi r_b (H+\delta H) \delta H)$ (see~\cite{Labousse13PoF} for experimental details). The resulting growth rate 
is indicated in Table \ref{Growth}c. 
Figure \ref{bump}c compares the experimental results from Labousse and Bush~\cite{Labousse13PoF} with the theoretical predictions. 
We only select the data corresponding to polygonal bumps in the absence of inner jumps. The theoretical model 
adequately describes the relatively sparse experimental data. 

\section{Conclusion}

We have developed a theoretical model with a view to rationalizing the instability of toroidal vortices with free surfaces. 
Two distinct regimes were identified, those dominated by the destabilizing influence of surface tension ($We\ll 1$) 
and inertia ($We\gg 1$). Provided the aspect ratio is sufficiently small  ($\chi \simeq 0.1$), our results 
are consistent with those of previous studies for a cylindrical configuration ~\cite{Ponstein59,Hocking59,Pedley1,Weidman06}.
The model predictions have been successfully applied to the toroidal Leidenfrost experiment, where the theoretical predictions 
are in good agreement with the experimental data \cite{Perrard}. Finally, the model predictions have been compared to 
existing experiments on the hydraulic jump and the hydraulic bump. As these configurations may exhibit more elaborate subsurface flow structures, our model is not likely to apply directly.
Nevertheless, our results do suggest that both vorticity and surface tension are likely to play an important role in this
class of polygonal instabilities.
\acknowledgments{The authors acknowledge the generous financial support of the National Science Foundation through grant number DMS-0907955. The authors are grateful to Jos\'{e} Bico, Marc Fermigier from the PMMH laboratory and `La soci\'{e}t\'{e} des amis de l'ESPCI ParisTech'. We also thank Yves Couder, St\'{e}phane Perrard and Laurent Limat from the MSC laboratory for the interesting discussions and the experiments which inspired this work We also thank Alexis Duchesne from MSC laboratory for his useful remarks about the hydraulic jump.}

\appendix
\section{Toroidal geometry~\label{appendixA}}
\subsection{Operators in toroidal geometry}
We recall the geometrical features of a torus in Table~\ref{Geometricalfeatures}.
\begin{table}[h!]
\begin{center}
\caption{\label{Geometricalfeatures}Geometrical features of a torus of radii $a$ and $R$ with $a<R$. $\rbar=r/a$ is the dimensionless distance to the core of the torus. The aspect ratio of the torus is denoted $\chi=a/R$. $\beta=1+\rbar\cos\theta$ corresponds to a dimensionless toroidal corrective term.}
\begin{small}
\begin{tabular}{c|c|c}
&elementary&integrated\cr
\hline
Surface&$dS=R^2 \beta d\varphi d\theta$&$ S=4 \pi ^2 R^2 \chi$\cr
volume&$d\tau =R^3 \beta d\varphi \overline{r} d\theta d\overline{r}$&$V=2\pi ^2 R^3 \chi^2$\cr
\end{tabular}
\label{geometrytorus}
\end{small}
\end{center}
\end{table}

The differential operators in toroidal coordinates are recalled in Table~\ref{differential}. If we consider $R\gg r$, we  recover formulae in cylindrical coordinates. If $R=0$ and $\theta \mapsto \pi /2-\theta$, we recover formulae in spherical coordinates. 
\begin{table}[h!]
\caption{Differential operators in a toroidal frame of radii $a$ and $R$ with $a<R$. $\rbar=r/a$ is the dimensionless distance to the core of the torus. The aspect ratio of the torus is denoted $\chi=a/R$. $\beta=1+\rbar\cos\theta $ corresponds to a dimensionless toroidal corrective term.}
\label{differential}
\begin{small}
\begin{center}
\begin{tabular}{l|c|l}
Displacement &$\mathbf{dl}$&$R\left( d\overline{r} \mathbf{e_r}+\overline{r}d\theta \mathbf{e_{\theta }}+ \beta d\varphi \mathbf{e_{\varphi }}\right)$\\
Gradient& $\bm{\nabla }f$&$ \dfrac{1}{R}\lbrace \dfrac{\partial f}{\partial \overline{r}}\mathbf{e_r}+\dfrac{1}{\overline{r}}\dfrac{\partial f}{\partial \theta}\mathbf{e_{\theta}}+\dfrac{1}{\beta}\dfrac{\partial f}{\partial \varphi }\mathbf{e_{\varphi}}\rbrace$\\
Divergence&$\bm{\nabla} \cdot \bm{f} $&$\dfrac{1}{R}\lbrace\dfrac{1}{\rbar\beta }\dfrac{\partial \rbar \beta f_r }{\partial \rbar}+\dfrac{1}{\rbar\beta}\dfrac{\partial \beta f_{\theta}}{\partial \theta}+\dfrac{1}{\beta}\dfrac{\partial f_{\varphi}}{\partial \varphi}\rbrace$
\end{tabular}
\end{center}
\end{small}
\end{table}
\subsection{Euler Equation in toroidal coordinates}
As toroidal coordinates are not commonly used, we detail here the derivation of the Euler equation (Eq.~\ref{Euler}a-c). The difference of momentum $\delta \bm{p}$ of an infinitesimal volume of incompressible fluid of mass $\rho \delta \mathcal{V}$ displaced by $d\bm{r}=dr\bm{e}_r+rd\theta \bm{e}_{\theta}+(R+r\cos\theta )d\varphi \bm{e}_{\varphi}$ during a time interval $dt$ is given by
\begin{equation}
\delta \bm{p}=\rho \delta \mathcal{V}\left[\bm{v}(t+dt,\bm{r}+d\bm{r})-\bm{v}(t,\bm{r}) \right] , 
\end{equation}
with the speed $\bm{v}(t,\bm{r})=u(t,\bm{r})\bm{e}_r+v(t,\bm{r})\bm{e}_{\theta}+ w(t,\bm{r}) \bm{e}_{\varphi}=\sum_i v_i \bm{e}_i$.  We have
\begin{equation}
\begin{array}{ll}
\delta \bm{p}&=\rho \delta\mathcal{V} dt\left[ \dfrac{\partial\bm{v}}{\partial t}+(\bm{v}.\bm{\nabla)}\bm{v}\right]\\
&=\rho \delta\mathcal{V} dt\left[ \sum_i\left(\dfrac{\partial v_i}{\partial t}\right)\bm{e}_i+(\bm{v}.\bm{\nabla)}\bm{v}+\sum_i \left(\dfrac{\partial \bm{e}_i }{\partial t}\right) v_i\right]\\
&=\rho \delta\mathcal{V} dt\left[ \sum_i\left(\dfrac{D v_i}{D t}\right)\bm{e}_i+\sum_i \left(\dfrac{\partial \bm{e}_i }{\partial t}\right) v_i\right]
\end{array}
\end{equation}
To compute every $\partial_t\bm{e}_i$, we recall that
\begin{equation}
\left\{
    \begin{array}{l} 
    \bm{e}_r=\cos\theta\cos\varphi \bm{e}_x+\cos\theta\sin\varphi \bm{e}_y+\sin\theta \bm{e}_z\\
    \bm{e}_{\theta}=-\sin\theta\cos\varphi \bm{e}_x-\sin\theta\sin\varphi \bm{e}_y+\cos\theta \bm{e}_z\\
     \bm{e}_{\varphi}=-\sin\varphi \bm{e}_x+\cos\varphi \bm{e}_y\\
    \end{array}
\right.
\end{equation}
which gives
\begin{equation}
\left\{
    \begin{array}{l} 
   \dfrac{\partial \bm{e}_r}{\partial t} =\dot{\theta}\bm{e}_{\theta}+\dot{\varphi}\cos\theta\bm{e}_{\varphi}\\
   \dfrac{\partial \bm{e}_{\theta}}{\partial t} =-\dot{\theta}\bm{e}_{r}-\dot{\varphi}\sin\theta\bm{e}_{\varphi}\\
    \dfrac{\partial \bm{e}_{\varphi}}{\partial t} =-\dot{\varphi}\left(\cos\theta \bm{e}_{r}-\sin\theta\bm{e}_{\varphi}\right)\\
    \end{array}
\right.
\end{equation}
or equivalently
\begin{equation}
\left\{
    \begin{array}{l} 
   \dfrac{\partial \bm{e}_r}{\partial t}u =\dfrac{uv}{r}\bm{e}_{\theta}+\dfrac{uw}{R+r\cos\theta }\cos\theta\bm{e}_{\varphi}\\
   \dfrac{\partial \bm{e}_{\theta}}{\partial t}v =-\dfrac{v^2}{r}\bm{e}_{r}-\dfrac{vw}{R+r\cos\theta }\sin\theta\bm{e}_{\varphi}\\
    \dfrac{\partial \bm{e}_{\varphi}}{\partial t}w =-\dfrac{w^2}{R+r\cos\theta}\left(\cos\theta \bm{e}_{r}-\sin\theta\bm{e}_{\varphi}\right)\\
    \end{array}
\right.
\end{equation}
Finally, the variation of momentum $\delta \bm{p}=\delta p_r\bm{e}_r+\delta p_{\theta}\bm{e}_{\theta}+\delta p_{\varphi}\bm{e}_{\varphi}$ can be written 
\begin{equation}
\left\{
    \begin{array}{ll} 
\delta p_r=\rho \delta\mathcal{V} dt\left(\dfrac{D u}{D t}  -\dfrac{1}{R}\left[ \dfrac{v^2}{\rbar}+\dfrac{\cos{\theta } w^2}{\beta}\right]\right) &(a) \\
       \delta p_{\theta}=\rho \delta\mathcal{V} dt\left( \dfrac{D v}{D t}  +\dfrac{1}{R}\left[ \dfrac{v\, u}{\rbar}+\dfrac{\sin{\theta } w^2}{\beta}\right]\right) &(b)\\
    \delta p_{\varphi}=\rho \delta\mathcal{V} dt\left(    \dfrac{D w}{D t}  +\dfrac{1}{R}\left[ \dfrac{w\, u \cos{\theta}}{\beta}-\dfrac{\sin{\theta } w\, v }{\beta}\right] \right)&(c)\\
         \end{array}
\right.
\label{Euler2}
\end{equation}
which justifies  equations~\ref{Euler}a-c.

\section{Derivation of Eq.~\ref{Equationpression}\label{Appendixsupplementaire}}
To obtain Eq.~\ref{Equationpression}, one must integrate the set of Eqs~\ref{Eulersteady}. We first consider Euler Equation (Eq.~\ref{Euler}). The speed is prescribed by Eq.~\ref{speedsteady}, specifically $\mathbf{v}=(u,v,w)=(0,\omega r/\beta,0)$. We thus find
\begin{small}
\begin{equation}
\left\{
    \begin{array}{ll} 
 - \dfrac{v^2}{\rbar} =-\dfrac{1}{\rho }\dfrac{\partial P}{\partial \rbar}+\dfrac{f_r}{\rho}&(a) \\
        0 =-\dfrac{1}{R\rho }\dfrac{1}{\rbar}\dfrac{\partial P}{\partial \theta}+\dfrac{f_{\theta}}{\rho }&(b)\\
        0 =-\dfrac{1}{\rho }\dfrac{1}{R\beta}\dfrac{\partial P}{\partial \varphi}+ \dfrac{f_{\varphi}}{\rho}&(c)
         \end{array}
\right.
\label{Euler3}
\end{equation}
\end{small}
We then expand the remaining inertial term at leading order in $\rbar$, yielding
\begin{equation}
\dfrac{v^2}{\rbar}=\dfrac{\omega^2 \rbar}{\left(1+\rbar \cos \theta\right)^2}R^2=\omega^2\rbar \left(1-2\rbar \cos \theta\right)R^2+\mathcal{O}(\rbar^3)
\end{equation} 
Integrating this term yields
\begin{equation}
\int\limits_{\rbar}^{\chi}d\rbar\; \dfrac{v^2}{\rbar}= \dfrac{\omega^2\left(\chi^2-\rbar^2\right)}{2}R^2+\mathcal{O}(\rbar^3)
\end{equation}
Using this expansion, the integration of Eq.~\ref{Euler3} with conditions ~\ref{conditionbodyforce} yields Eq.~\ref{Equationpression}.

\section{Derivation of Eq.~\ref{equationlinearise}\label{appendixB}}
We here derive Eq.~\ref{equationlinearise} by a first order expansion of $\bm{\varepsilon}=(\varepsilon_r/\beta,\varepsilon_{\theta}/\beta,\varepsilon_{\varphi})$ and $\tilde{p}$. 
The velocity field is
\begin{equation}
\begin{array}{ll}
\mathbf{V}_{\mathrm{total}}&=\omega\rbar \dfrac{R}{\beta} \mathbf{e}_{\theta}+\bm{\varepsilon}\\
&=\omega\rbar \dfrac{R}{\beta} \mathbf{e}_{\theta}+
\begin{pmatrix}
\dfrac{\varepsilon_{r}(\rbar)}{\beta } \\
\dfrac{\varepsilon_{\theta}(\rbar)}{\beta } \\
\varepsilon_{\varphi}(\rbar) \\
\end{pmatrix}\\
&=\omega\rbar \dfrac{R}{\beta} \mathbf{e}_{\theta}+
\begin{pmatrix}
\dfrac{\varepsilon_{r,0}(\rbar)}{\beta } \\
\dfrac{\varepsilon_{\theta ,0}(\rbar)}{\beta } \\
\varepsilon_{\varphi ,0}(\rbar) \\
\end{pmatrix} e^{\sigma t}e^{in\varphi}
\end{array}
\end{equation}
and the pressure distribution
\begin{equation}
 P_{\mathrm{total}}=P+\tilde{p}=P+\tilde{p}_0 e^{\sigma t}e^{in\varphi}
\end{equation}
We insert $\mathbf{V}_{\mathrm{total}}$ and $ P_{\mathrm{total}}$ in the Euler and continuity equations (Eq.~\ref{Euler}) and retain only the first order terms. 
The transport operator in the Euler equation (Eq.~\ref{Euler}) yields
\begin{equation}
\mathbf{V}_{\mathrm{total}}.\bm{\nabla}=\dfrac{1}{R\beta}\left(
\varepsilon_r\dfrac{\partial}{\partial \rbar}+
R\omega\dfrac{\partial}{\partial \theta}+\dfrac{\varepsilon_{\theta}}{\rbar}\dfrac{\partial}{\partial \theta}+
\varepsilon_{\varphi}\dfrac{\partial}{\partial \varphi}
\right)
\end{equation} 
In the following section we compute the first order term of the left hand-side of the Euler equation (Eq.~\ref{Euler}). We denote $f(\beta)=(1/\beta)\partial_{\theta}(1/\beta)$, $g(\rbar,\beta)=(1/\beta)\partial_{\rbar}(\rbar/\beta)$ and $(u,v,w)$ the components of the total velocity. 
Here we only retain the first order term in $\bm{\varepsilon}$. All the zeroth order terms are included in the steady state and denoted by 
$\mathit{O}(1)$ The higher order terms are denoted by the common notation  $\mathit{O}\left(\varepsilon^2\right)$.

\subsection{Derivation of Eq.~\ref{equationlinearise}-a }
Let us focus on the computation of
\begin{equation}
\dfrac{D u}{D t}  -\dfrac{1}{R}\left[ \dfrac{v^2}{\rbar}+\dfrac{\cos{\theta } w^2}{\beta}\right]
\end{equation}
We compute first 
\begin{equation}
\begin{array}{ll}
\left(\mathbf{V}_{\mathrm{total}}.\bm{\nabla}\right) \dfrac{\varepsilon_r}{\beta}&=\dfrac{1}{R\beta}\left(\omega R\dfrac{\partial \varepsilon_r/\beta}{\partial \theta}\right)+\mathit{O}(\varepsilon^2)\\
&= \omega\varepsilon_rf(\beta)+\mathit{O}(\varepsilon^2)
\end{array}
\end{equation}
Then we compute the cross terms
\begin{equation}
\begin{array}{ll}
-\dfrac{1}{R}\left[ \dfrac{v^2}{\rbar}+\dfrac{\cos{\theta } w^2}{\beta}\right]&=-\dfrac{1}{R\rbar}\left(\dfrac{R\omega\rbar}{\beta}+\dfrac{\varepsilon_{\theta}}{\beta} \right)^2+\mathit{O}\left(\varepsilon^2\right)\\
&=-2\dfrac{\omega \varepsilon_{\theta}}{\beta^2}+\mathit{O}(1)+\mathit{O}\left(\varepsilon^2\right)
\end{array}
\end{equation}
Finally we have
\begin{equation}
\begin{array}{ll}
\dfrac{D u}{D t}  -\dfrac{1}{R}\left[ \dfrac{v^2}{\rbar}+\dfrac{\cos{\theta } w^2}{\beta}\right]=&\dfrac{1}{\rho R}\left(A\varepsilon_r-B\varepsilon_{\theta}\right)\\
&+\mathit{O}(1)+\mathit{O}(\varepsilon^2)
\end{array}
\end{equation}
with $A=(\sigma/\beta+\omega f(\beta))\rho R$ and $B=(2\omega/\beta^2)\rho R$. For the particular angle $\theta=\pi$, we add a subscript to all the coefficients. For instance $\beta_{\pi}=(1-\rbar)$. We have $f(\beta_{\pi})=0$ which gives $A_{\pi}=(\sigma/\beta_{\pi})\rho R$ and $B_{\pi}=(2\omega/\beta_{\pi}^2)\rho R$.
\subsection{Derivation of Eq.~\ref{equationlinearise}-b}
In this subsection, we compute 
\begin{equation}
\dfrac{D v}{D t}  +\dfrac{1}{R}\left[ \dfrac{v\, u}{\rbar}+\dfrac{\sin{\theta } w^2}{\beta}\right] .
\end{equation}
We have to compute first
\begin{equation}
\begin{array}{ll}
\left(\mathbf{V}_{\mathrm{total}}.\bm{\nabla}\right) \dfrac{\varepsilon_{\theta}}{\beta}&=\dfrac{1}{R\beta}\left(\omega R\dfrac{\partial \varepsilon_{\theta}/\beta}{\partial \theta}\right)+\mathit{O}(\varepsilon^2)\\
&=\omega\varepsilon_{\theta}f(\beta)+\mathit{O}(\varepsilon^2) ,
\end{array}
\end{equation}
and then 
\begin{equation}
\begin{array}{rl}
\left(\mathbf{V}_{\mathrm{total}}.\bm{\nabla}\right) \dfrac{R\omega\rbar }{\beta}=&\dfrac{1}{R\beta}[\varepsilon_r\dfrac{\partial(R\omega\rbar /\beta)}{\partial \rbar} +\omega R\dfrac{\partial(R\omega\rbar /\beta)}{\partial \theta}\\
&+\dfrac{\varepsilon_{\theta}}{\rbar}\dfrac{\partial(R\omega\rbar /\beta)}{\partial \theta}]\\
=&\varepsilon_r\omega g(\rbar,\beta)+\varepsilon_{\theta}\omega f(\beta)+\mathit{O}(1).
\end{array}
\end{equation}
The cross terms yield
\begin{equation}
\begin{array}{ll}
\dfrac{1}{R}\left[ \dfrac{v\, u}{\rbar}+\dfrac{\sin{\theta } w^2}{\beta}\right]&=\dfrac{1}{R\rbar}\left(\dfrac{R\omega \rbar}{\beta}+\dfrac{\varepsilon_{\theta}}{\beta} \right)\dfrac{\varepsilon_r}{\beta}+\mathit{O}(\varepsilon^2)\\
&=\dfrac{\omega \varepsilon_r}{\beta^2}+\mathit{O}(\varepsilon^2) .
\end{array}
\end{equation}
Finally, we have 
\begin{equation}
\dfrac{D v}{D t}  +\dfrac{1}{R}\left[ \dfrac{v\, u}{\rbar}+\dfrac{\sin{\theta } w^2}{\beta}\right]=C\varepsilon_{\theta}+D\varepsilon_r+\mathit{O}(1)+\mathit{O}(\varepsilon^2)
\end{equation}
with $C=\sigma/\beta+2\omega f(\beta)$ and $D=\omega/\beta^2+\omega g(\rbar,\beta)$. At $\theta=\pi$, these coefficients become 
$C_{\pi}=\sigma/\beta_{\pi}$and $D_{\pi}={\pi}\omega/\beta_{\pi}^2+\omega g(\rbar,\beta_{\pi})$.
\subsection{Derivation of Eq.~\ref{equationlinearise}-c}
In this subsection, we compute
\begin{equation}
 \dfrac{D w}{D t}  +\dfrac{1}{R}\left[ \dfrac{w\, u \cos{\theta}}{\beta}-\dfrac{\sin{\theta } w\, v }{\beta}\right]
\end{equation}
First, we compute
\begin{equation}
\begin{array}{ll}
\left(\mathbf{V}_{\mathrm{total}}.\bm{\nabla}\right)\varepsilon_{\varphi}&=\dfrac{1}{R\beta}[
\varepsilon_r\dfrac{\partial \varepsilon_{\varphi}}{\partial \rbar}+
R\omega\dfrac{\partial \varepsilon_{\varphi}}{\partial \theta}+\dfrac{\varepsilon_{\theta}}{\rbar}\dfrac{\partial \varepsilon_{\varphi}}{\partial \theta}+
\varepsilon_{\varphi}\dfrac{\partial \varepsilon_{\varphi}}{\partial \varphi}]\\
&=\dfrac{\omega}{\beta}\dfrac{\partial \varepsilon_{\varphi}}{\partial \theta}+\mathit{O}(\varepsilon^2)\\
&=\mathit{O}(\varepsilon^2) ,
\end{array}
\end{equation}
then the cross terms
\begin{equation}
\begin{array}{ll}
\dfrac{1}{R}\left[ \dfrac{w\, u \cos{\theta}}{\beta}-\dfrac{\sin{\theta } w\, v }{\beta}\right]&=-\dfrac{\sin\theta}{R}\dfrac{w\, v }{\beta}+\mathit{O}(\varepsilon^2)\\
&=-\dfrac{\sin\theta}{R}\varepsilon_{\varphi} \left(\dfrac{R\omega\rbar}{\beta}+\dfrac{\varepsilon_{\varphi}}{\beta}\right)+\mathit{O}(\varepsilon^2)\\
&=-\dfrac{\sin\theta}{\beta}\omega \rbar \varepsilon_{\varphi}+\mathit{O}(\varepsilon^2).
\end{array}
\end{equation}
Finally we have
\begin{equation}
 \dfrac{D w}{D t}  +\dfrac{1}{R}\left[ \dfrac{w\, u \cos{\theta}}{\beta}-\dfrac{\sin{\theta } w\, v }{\beta}\right]=\dfrac{E}{\rho R \beta }\varepsilon_{\varphi}
\end{equation}
with $E=\rho R \beta (\sigma -\omega\rbar \sin\theta/\beta)$. Evaluated at $\theta=\pi$, we have $E_{\pi}=\rho R \beta_{\pi} \sigma$.
\section{Derivation of Eq.~\ref{equadiff} from Eq.~\ref{equationlinearise3}-d\label{appendixC}}
We here simplify equation \ref{equadiff} 
\begin{equation} \dfrac{1}{\rbar}\dfrac{\partial \left(\rbar F\dfrac{\partial \tilde{p}}{\partial \rbar}\right)}{\partial \rbar}-\dfrac{n^2}{E}\tilde{p}=0 .
\end{equation}
Let us recall that $F=C/(AC+BD)$. For the sake of simplicity and because the instability will be studied at $\theta=\pi$, 
we directly consider the equation
\begin{equation} \dfrac{1}{\rbar}\dfrac{\partial \left(\rbar F_{\pi}\dfrac{\partial \tilde{p}}{\partial \rbar}\right)}{\partial \rbar}-\dfrac{n^2}{E}\tilde{p}=0
\label{diffequationappendixC}
\end{equation}
with 
\begin{equation}
F_{\pi}=\dfrac{C_{\pi}}{A_{\pi}C_{\pi}+B_{\pi}D_{\pi}}=\dfrac{\beta_{\pi}/(\rho \sigma R)}{1+4\dfrac{\omega^2}{\sigma^2}\left[\dfrac{1}{\beta_{\pi}^2}\left(1+\dfrac{\rbar}{2\beta_{\pi}} \right)\right]} .
\end{equation}
A second order expansion in $\rbar$ gives
\begin{equation}
\dfrac{1}{\beta_{\pi}^2}\left(1+\dfrac{\rbar}{2\beta_{\pi}} \right)=1+\dfrac{5}{2}\rbar+\dfrac{9}{2}\rbar^2+\mathit{O}(\rbar^3)
\end{equation}
Note that this term, once evaluated at $\rbar=\chi$ gives the coefficient $K$ in Eqs~\ref{boundaryconditions} and \ref{equation}. We denote
\begin{equation}
F_0=\dfrac{1}{(\rho R\sigma )(1+4\dfrac{\omega^2}{\sigma^2})}
\end{equation} 
and expand $F$ in $\rbar$, yielding
\begin{equation}
F_{\pi}=F_{0,\pi}\mathcal{P}(\rbar)
\end{equation}
with the polynomial
\begin{equation}
\mathcal{P}(\rbar)=1-\rbar(1+5G)+\rbar^2(25G^2-4G) ,
\end{equation}
where 
\begin{equation}
G=\dfrac{1/2}{1+\dfrac{\sigma^2}{4\omega^2}}
\end{equation}
The differential equation~\ref{diffequationappendixC} yields
\begin{equation}
\rbar\dfrac{\partial}{\partial \rbar}\left(\rbar \mathcal{P}(\rbar)\dfrac{\partial \tilde{p}}{\partial \rbar}\right)-n^2\left( 1+4\dfrac{\omega^2}{\sigma^2}\right)\dfrac{\rbar^2}{1-\rbar}\tilde{p}=0
\label{equationdifffinale}
\end{equation}
Using a power series expansion $\tilde{p}=\sum_{n\in \mathbb{N}}\xi_n \rbar^n$, we can show that 
\begin{equation}
\tilde{p}=\xi_0\left(1+\dfrac{1}{4}(\tilde{n\rbar})^2\right)+\mathit{O}(\rbar^3)
\end{equation}
One recognizes the expansion of $I_0(\tilde{n}\rbar)$, the modified Bessel function of the first kind of order 0,
which leads to
\begin{equation}
\tilde{p}=\xi_0I_0(\tilde{n}\rbar)+\mathit{O}(\rbar^3).
\end{equation}
Let us recall that $I_0(\tilde{n}\rbar)$ satisfies the differential equation 
\begin{equation}
\rbar^2\dfrac{\partial^2 I_0}{\partial\rbar^2}+\rbar \dfrac{\partial I_0}{\partial\rbar}-\tilde{n}^2\rbar^2I_0=0
\end{equation} 
We conclude that for determining $\tilde{p}$ to second order in $\rbar$, one can replace the polynomial terms $\mathcal{P}(\rbar)$ and $(1-\rbar)$ by 1 in Eq.~\ref{equationdifffinale}. We thereby justify Eq.~\ref{equadiff}. 

\end{document}